\documentclass[12pt,prd,aps,superscriptaddress,preprintnumbers,amsmath,amssymb,nofootinbib]{revtex4-1}
\usepackage{amsmath,graphics,epsfig}
\usepackage{color,hyperref}
\usepackage{datetime}
\begin{document}

\title{Collider Signatures of Higgs-portal Scalar Dark Matter}
\author{Huayong Han}
\affiliation{Department of Physics, Chongqing University, Chongqing 401331, P. R. China}
\author{Jin Min Yang}
\affiliation{Institute of Theoretical Physics, Academia Sinica, Beijing 100190, P. R. China}
\affiliation{Department of Physics, Tohoku University, Sendai 980-8578, Japan}
\author{Yang Zhang}
\affiliation{Institute of Theoretical Physics, Academia Sinica, Beijing 100190, P. R. China}
\author{ Sibo Zheng}
\affiliation{Department of Physics, Chongqing University, Chongqing 401331, P. R. China}
\date{Jan 2016}

\begin{abstract}
In the simplest Higgs-portal scalar dark matter model,
the dark matter mass has been restricted to be either near the resonant mass ($m_h/2$)
or in a large-mass region by the direct detection at LHC Run 1 and LUX.
While the large-mass region below roughly 3 TeV can be probed by the future Xenon1T experiment,
most of the resonant mass region is beyond the scope of  Xenon1T.
In this paper, we study the direct detection of such scalar dark matter in the narrow resonant
mass region at the 14 TeV LHC and the future 100 TeV hadron collider.
We show the luminosities required for the $2\sigma$ exclusion and $5\sigma$ discovery.
\end{abstract}

\maketitle

\section{Introduction}
New physics beyond the Standard Model (SM) has drawn extensive attention since
the discovery of the SM Higgs boson \cite{1207.7214, 1207.7235}.
While a few problems such as how to stabilize the Higgs mass against ultraviolet radiative
corrections are tied to new physics of high mass scale,
in this paper we instead focus on dark matter with a mass near the weak scale.
In contrast to new physics which appears at a rather high mass scale,
such a dark matter model has promising prospect for discovery
at both astrophysical and particle collider experiments.

In particular, we are interested in the simplest Higgs-portal dark matter model,
in which the dark matter communicates with SM particles via the Higgs scalar.
Unlike the fermion dark matter setting,
a scalar dark matter in the so-called Higgs-portal scalar dark matter model (HSDM) \cite{Zee,0702143,0106249,0003350,0011335}
still survives the latest data of direct detections at Xenon100 \cite{1207.5988} and LUX \cite{1512.03506},
indirect detections at Fermi-LAT \cite{1503.02641,1506.00013},
and Higgs invisible decay at the LHC Run 1 \cite{LHC8ATLAS}.
Detailed discussions about this model have been given in the literature (\cite{0405097}-\cite{1409.6301}).
Fitting the experimental data indicates that
the dark matter mass is either near the resonant mass region between $53$ GeV and $62.5$ GeV
or in a large-mass region above $185$ GeV.

While the large-mass region between 185 GeV and 3 TeV can be probed by the future Xenon1T \cite{1206.6288},
most of the resonant mass region is beyond the reach of this facility.
In this paper, we discuss the collider signatures of the scalar dark matter in the HSDM model
with a mass between $53$ GeV and $62.5$ GeV at the 14 TeV LHC and the future 100 TeV
proton collider (FCC).
We will show that similar to Circular Electron Positron Collider (CEPC) \cite{CEPC,100pp},
FCC will be a useful machine for searching dark matter in this narrow mass region.
We will show that for FCC with a luminosity of 10 $ab^{-1}$
the exclusion and discovery sensitivities reach to 57 GeV and 56 GeV respectively
through the Vector Boson Fusion (VBF) channel,
and 54.8 GeV and 53.9 GeV respectively via the mono-$Z$ channel.
It indicates that FCC with 10 $ab^{-1}$ is a competitive facility in comparison with CEPC or Xenon1T.

The remaining parts of the paper are organized as follows.
In Sec. II, we briefly discuss the direct and indirect detection constraints on the HSDM.
In Sec.III we address the collider phenomenologies for the HSDM with dark matter mass
in the narrow resonant mass region at the 14 TeV LHC and the 100 TeV FCC,
where we focus on both the VBF channel and mono-$Z$ channel.
Our main results are presented in Sec. IV,
where we show the luminosities required for the $2\sigma$ exclusion and $5\sigma$ discovery.
Finally we conclude in Sec. V.

\section{Model and constraints}
\subsection{Model}
 In the simplest HSDM model, the dark matter $s$ communicates with the SM particles
through the SM Higgs scalar.
The Lagrangian for this mode reads as
\begin{eqnarray}{\label{Lag1}}
\mathcal{L}=\mathcal{L}_{\text{SM}}+\frac{1}{2}\left(\partial s\right)^{2} - \frac{\mu^{2}_{s}}{2}s^{2}  - \frac{\kappa_{s}}{2}s^{2}\mid H\mid^{2} - \frac{\lambda_{s}}{2}s^{4},
\end{eqnarray}
where $\mu_{s}$, $\lambda_{s}$ and $\kappa_s$ are the singlet scalar bare mass,
the self-interaction coupling constant, and the coupling constant between dark matter
and SM Higgs, respectively.
A $Z_2$ parity, under which $s$ is odd and other fields are even, is imposed to make the DM stable,
which reduces the number of model parameters.
After the electroweak symmetry breaking one can obtain
\begin{eqnarray}{\label{Lag2}}
\mathcal{L}=\mathcal{L}_{\text{SM}}+\frac{1}{2}\left(\partial s\right)^{2} - \frac{1}{2} m_{s}^{2}s^{2} -\frac{\kappa_{s}\upsilon}{2}s^{2}h - \frac{\kappa_{s}}{4}s^{2}h^{2} -\frac{\lambda_{s}}{2}s^{4},
\end{eqnarray}
where $m_{s}=\mu_{s}^{2}+\kappa_{s}\upsilon^{2}/2$ is the physical mass of the singlet scalar,
and $H=(\upsilon +h)/\sqrt{2}$,  $s=\left<s\right>+s$ and $\upsilon\simeq 246$ GeV.

Among the three model parameters,
the self-interaction coupling $\lambda_s$ does not directly affect the DM relic abundance,
the DM-nucleon scattering cross section and DM production cross section at colliders,
we simply decouple this parameter from the DM phenomenology discussed below.
It turns out that the remaining two parameters are strongly constrained.

\subsection{Constraints from indirect detections}
\begin{figure}[!h]
\includegraphics[width=0.7\textwidth]{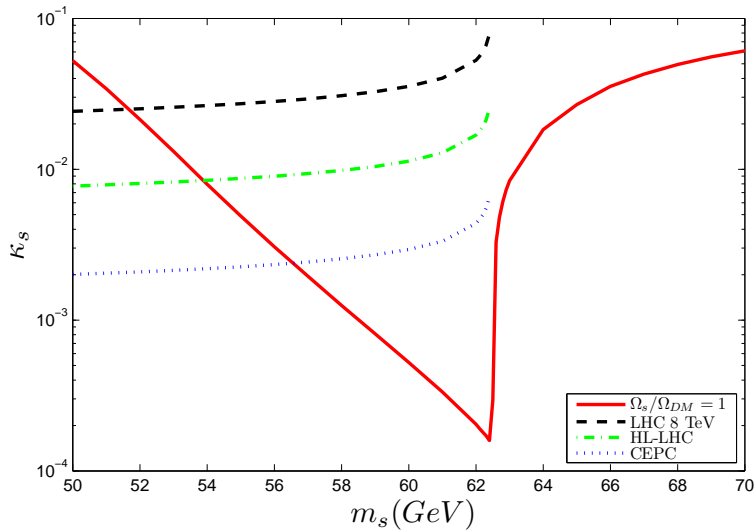}
\vspace{-0.5cm}
 \caption{Indirect constraints on the dark matter mass $m_s$ from the dark matter relic abundance
and Higgs invisible decays at the LHC Run 1, HL-LHC and CEPC.}
 \label{figIC}
\end{figure}

Assume that the cold dark matter is saturated by the singlet scalar $s$,
$s$ should account for the DM relic density measured by the Planck and WMAP \cite{1303.5076},
\begin{eqnarray}{\label{relic}}
\Omega_{\text{DM}}h^{2}=0.1199\pm 0.0027,
\end{eqnarray}
from which one infers the correlation between $m_s$ and $\kappa_s$ as
shown in Fig.\ref{figIC}.
Besides the relic abundance in Eq.(\ref{relic}),  there are other indirect constraints, including
the Higgs invisible decay $h\rightarrow ss$ in the mass region $m_{s}<m_{h}/2$
and the $\gamma$-ray limits from the Fermi-LAT \cite{1503.02641, 1506.00013}.
For the Higgs invisible decay, Fig.\ref{figIC} shows the latest limits at the 8 TeV LHC \cite{LHC8ATLAS}
, HL-LHC and CEPC \cite{CEPC},
which indicates that $m_s$ below 52 GeV is excluded by the data $\text{Br}(h\rightarrow ss)\leq 29\%$,
while the HL-LHC and CEPC can reach 54 GeV and 57 GeV, respectively.

\subsection{Constraints from direct detections}
\begin{figure}[!h]
\includegraphics[width=0.7\textwidth]{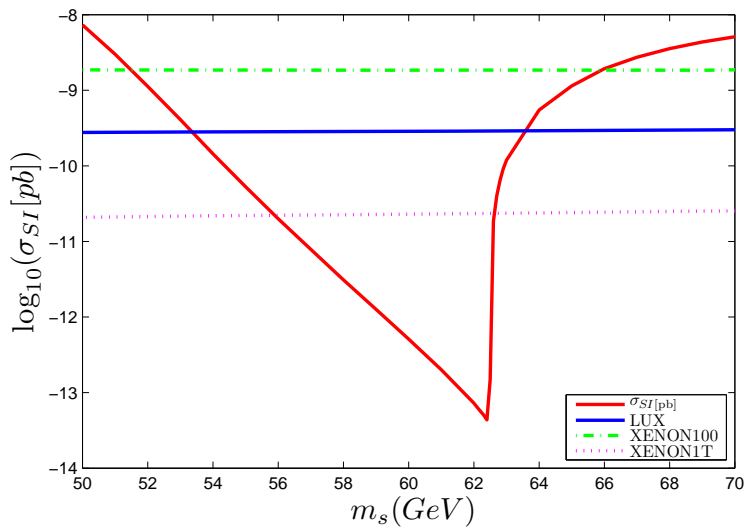}
\vspace{-0.5cm}
 \caption{ Direct-detection constraints on dark matter mass $m_s$ from LUX and Xenon experiments.
The red curve represents the dark matter relic abundance constraint.}
  \label{figDC}
\end{figure}

The direct detection at LUX and Xenon1T can further constrain the parameter space,
according to the spin-independent DM-nucleon scattering cross section given by
\begin{eqnarray}{\label{crosssection}}
\sigma_{\text{SI}}=\frac{\kappa_{s}^{2}f^{2}_{N}\mu^{2}m^{2}_{N}}{4\pi m^{4}_{h}m^{2}_{s}},
\end{eqnarray}
where $m_{N}$ is the nucleon mass,
$\mu=m_{s}m_{N}/(m_{s}+m_{N})$ is the DM-nucleon reduced mass,
and $f_{N}\sim 0.3$  is the hadron matrix element \cite{1306.4710}.
Fig.\ref{figDC} shows the predicted values of the spin-independent DM-nucleon scattering cross section,
together with direct detection limits at XENON100 \cite{1207.5988} and LUX \cite{1512.03506} experiments.
The limits at XENON1T \cite{1206.6288} have been also shown.
It indicates that the dark matter mass is restricted to a narrow resonant region
between 53 GeV and 63 GeV.
Once we employ the latest Fermi-LAT limits \cite{1508.04418},
this narrow mass region is further reduced to a narrow range between 53 GeV and 62.5 GeV.

\section{Dark Matter at Hadron Colliders}
In this section we study the collider signatures of the scalar dark matter at the 14 TeV LHC and 100 TeV FCC.
We will explore the sensitivities at these two colliders for the dark matter mass in the narrow resonant
region between 53 GeV and 62.5 GeV.
We consider the dominant VBF channel as well as the sub-leading but relatively clean mon-$Z$ channel.

We use FeynRules \cite{1310.1921} to generate model files prepared for MadGraph5 \cite{1405.0301},
which includes Pythia 6 \cite{0603175} for parton showering and hadronazition,
and the package Delphes 3 \cite{1307.6346} for fast detector simulation.
In particular, the default CMS detector card and the Snowmass detector card are used for the 14 TeV
and 100 TeV $pp$ collider, respectively.
Events are generated by using Madgraph5 at the leading order with the 125 GeV Higgs and
fixed value $\kappa_{s}=1.0$ for different dark matter masses.
Cross sections are reproduced by rescaling $\kappa_{s}^2$ which corresponds to $m_{s}$.

\subsection{Vector boson fusion}
In the VBF channel, the dark matter pairs are produced through the Higgs scalar
\begin{equation}
p p \to h + j j \to s s + jj,
\end{equation}
where the Higgs $h$ should be on-shell in our case.
The primary SM backgrounds to this process include
$Z+$jets, $W+$jets, $t\bar{t}+$jets and QCD multi-jets.
For simplicity we consider the main contributions arising from $Z+{\rm jets}$ and $W+$jets channels,
and adopt the cuts used by the CMS VBF analysis \cite{1404.1344} for event selection:
\begin{eqnarray}{\label{selection1}}
{p_{T}}_{j_{1(2)}} > 50 ~{\rm GeV},  &\,& ~ |\eta_{j_{1(2)}}| < 4.7, \nonumber \\
\eta_{j_1} \cdot \eta_{j_2} < 0, &\,&~ \Delta \eta_{jj} > 4.2, \nonumber \\
M_{jj} > 1100~ {\rm GeV}, &\,&~ \Delta \phi_{jj}  < 1.0  , \nonumber\\
E_{T}^{\rm miss} > 130~ {\rm GeV},
\end{eqnarray}
where ${p_{T}}_{j_{1(2)}}$ and $\eta_{j_{1(2)}}$ are the transverse momentum
and pseudo-rapidity of the first (second) leading jet, respectively.
$\Delta \eta_{jj}$, $\delta \phi_{jj}$ and $M_{jj}$ are the rapidity difference,
azimuthal-angel difference and invariant mass of the two leading jets, respectively.
Any event with an additional jet with $p_T > 30 $ GeV and pseudo-rapidity between those
of the two tagged jets is rejected.

We firstly apply the criteria in Eq.(\ref{selection1}) to the 8 TeV LHC with data of 19.5 $fb^{-1}$.
The number of events for the SM background is 134 in the $Z+$jets channels and 145 in the $W+$jets
channels, respectively.
Compared with the reported number of events (99 in the $Z+$ jets channels and 183 in the $W+$ jets channels)
by the CMS collaboration \cite{1404.1344}, they are consistent with each other.


The criteria in Eq.(\ref{selection1}) will be also applied to both the 14 TeV LHC and 100 TeV FCC simulations for conservation. It is based on the following facts. First, there is little difference between the 8 TeV LHC and 14 TeV LHC except the collision energy, which means the cut on the pseudo-rapidity of the first two leading jets should remain unchanged. Second, the kinetic distribution of the signal events though on-shell Higgs decay and the main backgrounds Z$+$jets and W$+$jets have similar changing trends when one modifies the cuts in Eq.(\ref{selection1}), as the mass difference between the Higgs and $Z(W)$ boson can be omitted compared to the variation of collision energy. 
It turns out that for a benchmark DM mass the effects on the ratio of signal over background events $S/B$ are less than two times due to variations on the cuts in Eq.(6) such as $p_{T_{j_{1(2)}}}>\{40,~50,~60,~80,~100\}$ GeV,  $\Delta \eta_{jj}>\{4,~4.2,~4.5\}$, $M_{jj}>\{900,~1100,~1300,1500\}$ GeV, $\Delta\phi_{jj}<\{0.8,~1.0,~1.2\}$  and $E_T^{miss} > \{100,~120,~150,~180\}$ GeV.
See table below for details.

 \begin{table}
\begin{center}
\begin{tabular}{|c|c|c|}
  \hline
 $p_{T_{j1(2)}}>\{40, 50, 60, 80,100\}$ GeV &  $\Delta\eta_{jj} >\{4.0, 4.2, 4.5\}$ & 
$M_{jj} >\{900, 1100, 1300, 1500\}$ GeV    \\
  \hline
  $\{1, 1, 1.02, 0.97, 0.86\}$ & $\{0.89, 1, 1.17\}$  &  $\{0.82, 1, 1.19, 1.38\}$\\
\hline 
 \end{tabular}
\begin{tabular}{|c|c|}
 \hline
 $\Delta\phi_{jj}<\{0.8, 1, 1.2\}$  &$E_{T}^{miss}>\{100, 120, 130, 150, 180\}$ GeV \\
\hline
$\{1.02, 1, 0.97\}$ & $\{0.84, 0.95, 1, 1.09, 1.18\}$  \\
\hline
\end{tabular}
\caption{Effects on the ratio $S/B$ due to variations on the cuts in Eq.(\ref{selection1}) for benchmark DM mass $m_{s}= 53$ GeV at 100 TeV FCC.}
\end{center}
\end{table}

\subsection{Mono-$Z$ channel}
In the mono-$Z$ channel the dark matter pairs are produced via the process
\begin{equation}
p p \to h + Z \to s s + Z.
\end{equation}
Compared with the VBF channel,
the mono-$Z$ channel is sub-leading but relatively cleaner.
For event selection in this channel
we adopt the following cuts as suggested by the CMS leptonic mode analysis \cite{1511.09375}:
\begin{eqnarray}{\label{selection2}}
p_{T}^l > 20 ~{\rm GeV},\,  &\,&~ |\eta_{e(\mu)}| < 2.5(2.4), \nonumber \\
|m_{ll} - m_{Z}| < 10 ~{\rm GeV},\, &\,&~ E_{T}^{\rm miss} > 80~ {\rm GeV}, \nonumber \\
p_{T}^{ll} > 50~ {\rm GeV}, \, &\,&~ |u_{\|}/p_{T}^{ll}| < 1.0, \nonumber \\
\Delta \phi_{ll, \overrightarrow{p}_{T}^{\rm miss}} > 2.7, \, &\,&~ |E_{T}^{\rm miss} - p_{T}^{ll} |/p_{T}^{ll} < 0.2,
\end{eqnarray}
where $p_{T}^{ll}$ is the dilepton transverse momentum and $u_{\|}$ is defined as the component of
$\overrightarrow{u} = - \overrightarrow{p}_{T}^{\rm miss} - \overrightarrow{p}_{T}^{ll}$ parallel to the
direction of $\overrightarrow{p}_{T}^{ll}$. Events are rejected if an additional electron or muon is
reconstructed with $p_T > 10$ GeV, and any event having two or more jets with $p_T > 30$ GeV.

Similar to the discussions in the preceding section,  the criteria in Eq.(\ref{selection2}) are
examined via the 8 TeV LHC simulation with data of 19.7 $fb^{-1}$.
The number of events for the SM background is 88 in the mono-$Z$ channel.
Compared with the reported number of events 138 in the same channel by the CMS collaboration
\cite{1511.09375}, they are consistent with each other.
Unlike the CMS result, the next-leading order (NLO) effects have been neglected in our simulation.
The two numbers of events likely match better if the NLO effects are included in our analysis.
Following the similar facts mentioned in the preceding section,
the criteria in Eq.(\ref{selection2}) will be directly applied to both the 14 TeV LHC and 100 TeV
 FCC simulations.

\section{Results}
\begin{figure}[!h]
\includegraphics[width=0.7\textwidth]{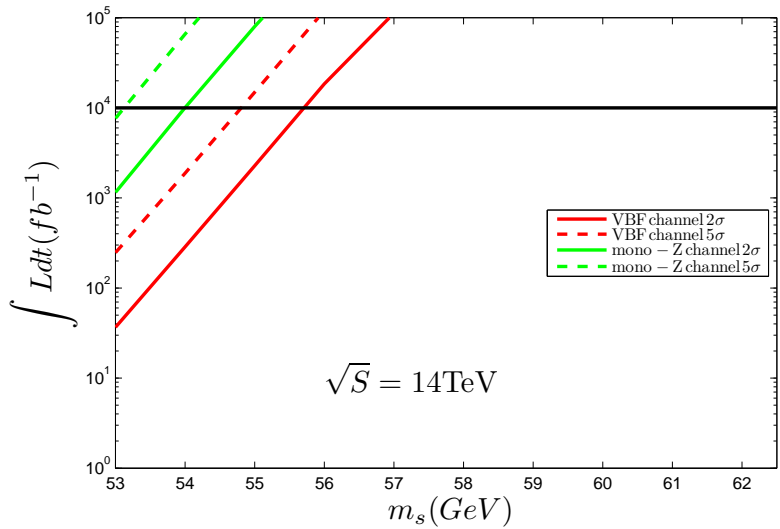}
\vspace{-0.5cm}
 \caption{The integrated luminosity needed for the exclusion determined by $S/\sqrt{B}=1.96$ and
5$\sigma$ discovery determined by $S/\sqrt{S+B} = 5 $ in VBF and mono-$Z$ channel at
the 14 TeV LHC, respectively.
The solid dark line represents the 10 ab$^{-1}$ integrated luminosity.}
\label{fig14}
\end{figure}

\begin{figure}[!h]
\includegraphics[width=0.7\textwidth]{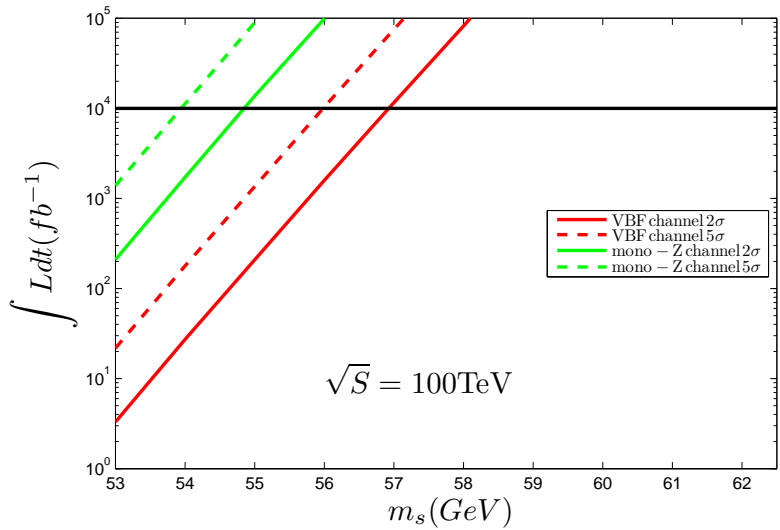}
\vspace{-0.5cm}
 \caption{Same as Fig.\ref{fig14}, but for the 100 TeV FCC.}
  \label{fig100}
\end{figure}

We present our main results in Fig.\ref{fig14} and Fig.\ref{fig100},
which show the integrated luminosity $\mathcal{L}$ needed for exclusion and discovery
at the 14 TeV LHC and 100 TeV FCC, respectively.
Here, we take the following definition about significance
\begin{eqnarray}{\label{sig}}
  \frac{S}{\sqrt{B}}~~ ({\rm for~ exclusion}),~~~~~
  \frac{S}{\sqrt{S+B}}~~ ({\rm for ~discovery}).
\end{eqnarray}
Systematic uncertainties are neglected in both the signal and the background simulations.

In Fig.\ref{fig14} one observes that for $\mathcal{L}=10^{2}$ $fb^{-1}$
the scalar dark matter mass up to 53.5 GeV can be excluded,
and for $\mathcal{L}=10^{3}$ $fb^{-1}$  the exclusion and discovery limits
via  VBF channel will reach to 54.6 GeV and 54 GeV, respectively.
In contrast, Fig.\ref{fig100} shows that
for $\mathcal{L}=10^{2}$ $fb^{-1}$
the scalar dark matter mass up to 54.5 GeV can be excluded,
and for $\mathcal{L}=10^{3}$ $fb^{-1}$
the exclusion and discovery limits at the FCC
via VBF channel approaches to 55.8 GeV and 55 GeV, respectively.
The exclusion limits via the mono-$Z$ channel are obviously weaker in comparison with the VBF channel.

Taking an integrated luminosity $\mathcal{L}=10$ $ab^{-1}$ at the 100 TeV FCC, one finds that
the exclusion and discovery limits approach to 57 GeV and 56 GeV in the VBF channel, respectively.
Compared with either CEPC or Xenon1T, where
the exclusion limits approach to 56.5 GeV and  56 GeV, respectively,
the FCC with $\mathcal{L}=10$ $ab^{-1}$ is a competitive facility.
Fig.\ref{fig14} and Fig.\ref{fig100} also illustrate that
it is unlikely to detect the scalar dark matter in the mass range between 57 GeV and 62.5 GeV
in HSDM model at any present and future facilities mentioned in this paper.

\section{Conclusion}
In this paper,  we explored the collider signatures of the scalar dark matter in the HSDM model.
Our study shows that for the 100 TeV FCC with an integrated luminosity of 10 $ab^{-1}$,
the exclusion and discovery sensitivities reach to 57 GeV and 56 GeV respectively
through the VBF channel,
and 54.8 GeV and 53.9 GeV respectively via the mono-$Z$ channel.
Compared with either CEPC or Xenon1T, where
the exclusion limits approach to 56.5 GeV and  56 GeV, respectively,
FCC is a competitive facility.
Unfortunately, the scalar dark matter in the mass range between 56.5 GeV and 62.5 GeV
is unlikely to be either directly or indirectly detected at any present and future facility
discussed in this paper.

\begin{acknowledgments}
This work is supported in part
by the National Natural
Science Foundation of China under grant Nos. 11275245, 11135003 and 11405015,
and by the CAS Center for Excellence in Particle Physics (CCEPP).

\end{acknowledgments}

\end{document}